Coherent Control of Quantum States by Quadratically Chirped Short Laser Pulses

G.P. Djotyan[1], A.A. Avetisyan[2], A.P. Djotyan[2]

[1]HUN-REN Wigner Research Centre for Physics, Budapest, Hungary
[2]Yerevan State University, Yerevan, Armenia

e-mail   artakav@ysu.am

## Abstract

We present a novel scheme for coherent manipulation of populations and robust creation of arbitrary coherent superposition of metastable states of a quantum system with lambda-configuration of operating energy levels using laser pulses with a quadratically chirped carrier frequency. The case of a single "broadband" laser pulse is considered, when the frequency spectrum of the pulse (without chirp) exceeds the frequency distance between the metastable energy levels of the system. The results of the dressed state analysis and numerical simulation are presented, demonstrating reliable and robust creation of an arbitrary coherent superposition of metastable states by changing the parameters of laser pulses.

### *I.     Introduction*

Coherent control of quantum states finds important applications in numerous fields of science and technology. Among them are atomic interferometry and atomic beam splitters [1-5], quantum chemistry [6-8], quantum information and data processing [8-14], quantum optics and manipulation of quantum states of multilevel quantum systems [15-27], generation of high harmonics and improving efficiency of nonlinear processes in resonant gases [28–33], laser cooling [34], Bose–Einstein condensation [35], writing and storage of optical phase information [36,37], and other fields, see also the review papers [38-43].

Different schemes of coherent population transfer and coherent creation of superposition states have been investigated extensively in recent years [15–27]. In [23] a single laser pulse with linear frequency chirp (FC) in the adiabatic following regime was used to create coherent superposition of metastable states in an atomic system with tripod- configuration of operating energy levels. Note that FC laser pulses were used for coherent control of quantum states of the graphene [44], quantum dots [44a] and of lowest states of a shallow Impurity in graphene monolayer [46].   Effect of dissipation processes on coherent population transfer and coherence creation between states of quantum systems also was considered; see [47-50].

In Ref.[51] a quadratically frequency chirped (QFC) laser pulse was applied to demonstrate a robust  population transfer between the states of a two-level quantum system with the time of the population transfer significantly shorter than in the case of a linearly FC laser pulse.

In this work, we apply a short QFC laser pulse to a quantum system of $\Lambda$-configuration of working energy levels to create a coherent superposition of metastable states of the system with a negligible and short-term excitation of the system.

Based on dressed states analysis and numerical simulations, we show that the absolute value of the generated coherence can be controlled by changing the center frequency and amplitude of the QFC laser pulse.

The physics of the process is as follows: due to the quadratic chirp, the laser pulse enters into resonance with the quantum system twice instead of one resonance in the case of linear-frequency chirped pulses.

In the case of a two-level atom initially in the ground state, the atom returns to its original state after adiabatic excitation following interaction with the first branch of the parabolic frequency chirp and is de-excited after interaction with the second branch of the parabolic frequency chirp.

In the case of a Λ-structured atom with a "broadband" laser pulse interacting with both allowed transitions simultaneously, only the "bright" superposition component of the metastable states is excited and de-excited as a result of interaction with the QFC pulse, while the "dark" component of the superposition remains unaffected.

While the amplitude of the "bright" component at the end of the interaction is the same as before the interaction (similar to the case of the considered above two-level atom), the phase of the "bright" component after the interaction depends on the parameters of the laser pulse.

Since the values of the amplitudes of metastable states are equal to a linear combination of the "bright" and "dark" components, the final probability amplitudes, and therefore the coherence between them, depend on the parameters of the laser pulse, including its central frequency, the amplitude of the Rabi frequency, and the chirp speed.

By changing these parameters, it is possible to obtain a given value of coherence between metastable states with a negligible and temporary excitation of the system, avoiding the effects of spontaneous relaxation and dephasing.

As the analysis below shows, the resulting populations of metastable states and the generated coherence between these states depend on the laser pulse parameters, even if the QFC pulse frequency does not pass through resonance but has a turning point near resonance.

In this paper, two different cases are studied: when the time-dependent frequency of the QFC laser pulse resonates (twice) with the transition between the metastable and excited states, and when the time-dependent QFC laser pulse frequency has a turning point near the resonance but does not resonate with the transition.

## *2. Mathematical Formalism*

We consider interaction of a phase modulated (frequency chirped) laser pulse with a quantum system having a $\Lambda$ – configuration of the working levels, (see Fig.1). It is assumed that the laser pulse is "broadband" with a frequency spectrum (without chirp) exceeding the frequency distance between two metastable states of the system, which allows the pulse to interact simultaneously with both allowed transitions of the Λ-system.

An important feature of our consideration is the assumption of quadratic chirping of the laser pulse. Assuming the pulse duration to be shorter than all relaxation times of the system, we use the Schrödinger equation to describe the interaction of the laser pulse with the system. The Schrödinger equation for the column state-vector

$\underline{a} = (a_1, a_2, a_3)^T = a_1 (1 \ 0 \ 0)^T + a_2 (0 \ 1 \ 0)^T + a_3 (0 \ 0 \ 1)^T$ has the following form:

$$\frac{d}{dt}\underline{a} = i\hat{H}\underline{a}, \qquad (1)$$

where the column vectors $(1 \ 0 \ 0)^T$, $(0 \ 1 \ 0)^T$, and $(0 \ 0 \ 1)^T$ stand for the (bare) states $|1\rangle$, $|2\rangle$, and $|3\rangle$ respectively, (see Fig.1). The Hamiltonian in the rotating-wave approximation is, [23]:

$$\hat{H} = \begin{bmatrix} 0 & \Omega_{12} & 0 \\ \Omega_{21} & \varepsilon_{21} + t\frac{d}{dt}\varepsilon_{21} & \Omega_{23} \\ 0 & \Omega_{32} & \omega_R \end{bmatrix}, \qquad (2)$$

where $\Omega_{ij} = \Omega_{ji}^* = \left(\frac{1}{2\hbar}\right) d_{ij} A(t)$, (i, j= 1,2,3) is the Rabi frequency and $d_{ij}$ is the dipole moment matrix element for laser-induced transition from state $|j\rangle$ to state $|i\rangle$.

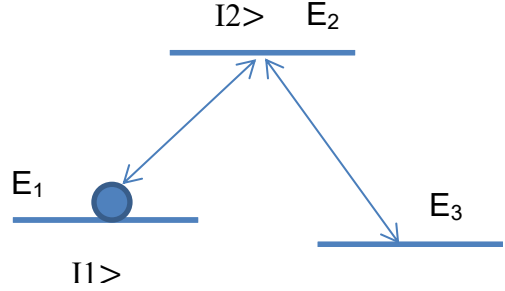

Fig.1. The scheme of the $\Lambda$-atom levels.

Note that the transition between the metastable states I1> and I3> is forbidden, and only transitions between the states I1>, I3> and the excited I2> state are allowed in the dipole approximation. $A(t)$ is the (real) envelope of the laser pulse $A(t) = A_0 \exp(-t^2/2\tau_p^2)$ assumed of Gaussian shape, where $A_0$ is its amplitude and $2\tau_p$ is duration of the pulse (intensity). $\varepsilon_{21} = \omega_L(t) - \omega_{21}$, and $\varepsilon_{23} = \omega_L(t) - \omega_{23}$ are the detuning from the one-photon resonances, where $\omega_L(t)$ is the time-dependent carrier frequency of the laser pulse; $\omega_{21}$ and $\omega_{23}$ are the resonance transition frequencies between the corresponding states. Note that in the case of a single laser pulse interacting with the quantum system under consideration, the Raman detuning $\omega_R = \varepsilon_{21}(t) - \varepsilon_{23}(t) = \omega_{13}$ is a time independent parameter equal to $\omega_{13} = (E_1 - E_3)/\hbar$ being the angular frequency interval between the two ground states of the quantum system with energies $E_1$ and $E_3$, (see Fig.1). By assuming that the width $\Delta\omega_p$ ($\Delta\omega_p \sim 1/\tau_p$) of the frequency spectrum of the "broadband" laser pulse exceeds the frequency interval $\omega_{13}$ between the two metastable states ($\Delta\omega_p > \omega_{13}$), we can assume Raman resonance for interaction of the "broadband" laser pulse with the quantum system under consideration, and put the Raman detuning $\omega_R = 0$ in the Hamiltonian in Eq.(2). Note that the situation in this case is analogous to the interaction of the quantum system with a bichromatic frequency-chirped laser pulse, consisting of two laser pulses with identical envelopes and identical frequency chirps and the carrier frequencies shifted by the frequency distance between the two ground states (see also Refs [22,36,37]).

In what follows we assume a quadratic chirp in time for the laser carrier frequency: $\omega_L(t) = \omega_{L0}(t) + \beta t^2$, where $\omega_{L0}$ is the central frequency and $\beta$ is the chirp speed parameter.

At this step it is convenient to introduce the "bright" $g_b$ and "dark" (or "trapped") $g_d$ superposition of the (bare) probability amplitudes of the ground states of the $\Lambda$ - system:

$$g_b(t) = [a_1(t)d_{21} + a_3(t)d_{32}]/d,$$

$$g_d(t) = [a_1(t)d_{12} - a_3(t)d_{23}]/d, \qquad (3)$$

where $d = \sqrt{|d_{12}|^2 + |d_{23}|^2}$.

From the Schrödinger equation (1), we obtain the following equations for the amplitudes of the excited $a_2$, as well as the "bright" $g_b$ and "dark" $g_d$ superposition states:

$$\frac{d}{dt}a_2 - i\epsilon(t)a_2 = i\, F(t)\, g_b,$$

$$\frac{d}{dt}g_b = i\, F(t)\, a_2,$$

$$\frac{d}{dt}g_d = 0, \tag{4}$$

where $\epsilon(t) = \epsilon_{21} + t\frac{d}{dt}\epsilon_{21}$, and $F(t) = \frac{1}{2\hbar}A(t)\,d$.

As it follows from Eq.(4), the amplitude of the "dark" superposition state remains equal to its initial value before interaction with the laser pulse: $g_d(t) \equiv g_d(-\infty)$. On the other hand, the equations for the amplitudes of the "bright" component and the excited state correspond to an equivalent two-level quantum system with probability amplitudes of the ground and excited states equal to $g_b(t)$ and $a_2(t)$, respectively.

### 2.1. The dressed - states picture

In this section, we proceed to the analysis of the problem of coherent control of the states of the equivalent two-level quantum system under consideration (see equation (4)) using the dressed state approach.

The solution of Eq. (4) can be represented on the basis of the adiabatic dressed states $\underline{b}^{(k)}$ as follows:

$$\underline{c}(t) = \sum_{k=1}^{2} r_k(t)\underline{b}^{(k)}(t)\, \exp[-i\int_{-\infty}^{t} w_k(t')dt'], \tag{5}$$

with the initial condition at $t \to -\infty$,

$$\underline{c}(t \to -\infty) = \sum_{k=1}^{2} r_k(-\infty)\underline{b}^{(k)}(-\infty), \tag{6}$$

where the column state-vector $\underline{c}(t) = (\, g_b(t)\; a_2(t)\, )^T$ is solution of the Schrödinger equation (4), $\underline{b}^{(k)}(t)$ is the adiabatic dressed-state eigenvector, corresponding to the eigenvalue (quasi-energy) $w_k$ of the Hamiltonian $\widehat{H}_b$:

$$\widehat{H}_b\, \underline{b}^{(k)}(t) = w_k(t)\, \underline{b}^{(k)}(t), \tag{7}$$

where

$$\widehat{H}_b = \begin{pmatrix} \epsilon_{21} + t\frac{d}{dt}\epsilon_{21} & F(t) \\ F(t) & 0 \end{pmatrix}, \tag{8}$$

from Eq.(4).

The function $r_k(t)$ in Eq.(5) is the statistical weight of the adiabatic dressed state-vector $\underline{b}^{(k)}(t)$ in the bare state-vector $\underline{c}(t)$. According to the adiabatic theorem [52,53], $r_k(t) \equiv$ $r_k(-\infty)$ if the change in time-dependent parameters of the system is adiabatically slow, (see the relevant conditions below).

We obtain the following expressions for the quasi-energies $w_k(t)$ and for the dressed-state eigenvectors $\underline{b}^{(k)}(t)$ as solutions of Eq.(7):

$$w_{1,2} = \frac{\epsilon(t)}{2} \pm \sqrt{\frac{\epsilon(t)^2}{4} + F(t)^2} , \qquad (9)$$

$$\underline{b}^{(k)}(t) = b_1^{(k)} (1 \quad 0)^T + b_2^{(k)} (0 \quad 1)^T , \qquad (10)$$

$$b_1^{(k)}(t) = \frac{F(t)}{\sqrt{[w_k - \epsilon(t)]^2 + F(t)^2}} g_b(-\infty),$$

$$b_2^{(k)}(t) = \frac{w_k - \epsilon(t)}{\sqrt{[w_k - \epsilon(t)]^2 + F(t)^2}} g_b(-\infty), \qquad (11)$$

where $\epsilon(t) = \epsilon_{21} + t \frac{d}{dt} \epsilon_{21} = e_{0p} + 3\beta t^2$ and $g_b(-\infty)$ is the amplitude of the "bright" superposition component before interaction with the laser pulse. In what follows, we assume equal transition dipole elements for the dipole allowed transitions of the system: $d_{12} = d_{32}$. $F(t) = \frac{1}{2\hbar} A(t) \sqrt{|d_{12}|^2 + |d_{23}|^2} = w_p \sqrt{2} \exp[-t^2/2\tau_p^2]$, with $w_p = \frac{|d_{12}|}{2\hbar} A_0$. The dependence of quasienergies $w_1(t)$ and $w_2(t)$ on time is shown in Fig.2.

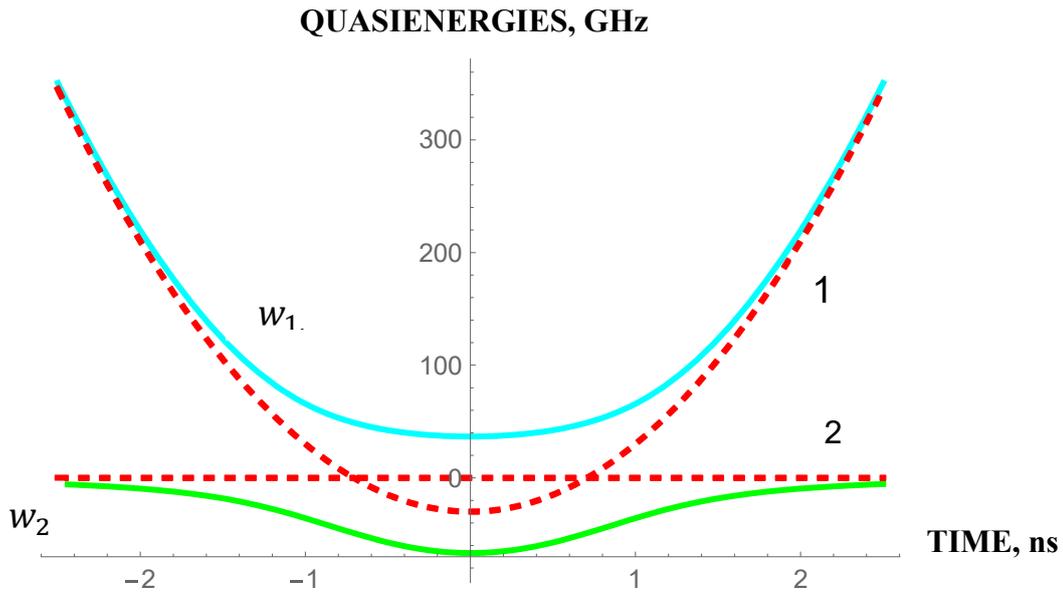

Fig.2. Time dependence of the states quasi-energies $w_1$, (blue) and $w_2$, (green), normalized by the Planck's constant $\hbar$, as well as the quasi-energies in the absence of a laser field (dashed

lines). A quadratic chirp of the laser carrier frequency is assumed: $\epsilon(t) = e_{0p} + 3\beta t^2$. Parameters applied are as follows: half pulse duration $\tau_p$ = 5 ns, $e_{0p} = -30$ GHz, $\beta = 15$ GHz$^2$, $w_p$ = 35 GHz.

Let us assume at this point that the system is in one of the ground state initially ($t \to -\infty$). It means that the state vector $\underline{c}(t \to -\infty) = (g_b \ 0)^T$, and in the dressed states approach,

$$\underline{c}(t \to -\infty) = \sum_{k=1}^{2} r_k(-\infty)\underline{b}^{(k)}(-\infty) = \sum_{k=1}^{2} r_k(-\infty)[b_1^{(k)}(-\infty) \ (1 \ 0)^T].$$

As it follows from Eqs.(9-11), only for $w = w_1 = \frac{\epsilon(t)}{2} + \sqrt{\frac{\epsilon(t)^2}{4} + F(t)^2}$, $b_1^{(1)} \to g_b(-\infty)$, and $b_2^{(1)} \to 0$ at $t \to -\infty$. So, the dressed state $\underline{b}^{(1)}(t) = b_1^{(1)}(t) (1 \ 0)^T + b_2^{(1)}(t) (0 \ 1)^T$ is the dressed state vector, which must be identified with the one that is adiabatically transformed into the initial state vector of the system before interaction with the laser pulse (at $t \to -\infty$), (see Eq.(6)): $\underline{c}(t \to -\infty) = \underline{b}^{(1)}(t \to -\infty)$. Accordind to the adiabatic theorem, the system remains in this state if the conditions of adiabaticity are met (see [51,52]).
The general condition for realization of the adiabaticity condition in the case of two level system is (see [42]):

$$|d\theta/dt| \ll \sqrt{\epsilon^2(t) + F(t)^2}, \quad (12)$$

where the mixing angle

$$\theta = \tan^{-1}\left[\frac{\epsilon(t)}{F(t)}\right].$$

Assuming that $F(t)^2 \gg \epsilon^2(t)$ during the states populations transition time $t_{tr}$ in the system, we arrive to the following condition from Eq.(12), (see also [42]):

$$F_0^2 \gg \frac{d}{dt}\epsilon(t),$$

where $F_0$ is maximum Rabi frequency (coupling strength): $F_0 = \frac{|d_{12}|}{\sqrt{2}\hbar} A_0$. For our case of quadratically chirped laser pulse, we will have the following adabaticity condition:

$$F_0^2 \gg 6\beta t_{tr}, \quad (13)$$

where $t_{tr}$ is the populations transition time in the system that is function of the chirp speed parameter $\beta$ and other parameters of the laser field.
Below in Fig. 3 the dynamics of the population of states for the quadratic chirp of the laser pulse frequency is presented as a result of numerical simulation of equation (4).

**STATES POPULATIONS, LASER PULSE SHAPE, DETUNING**

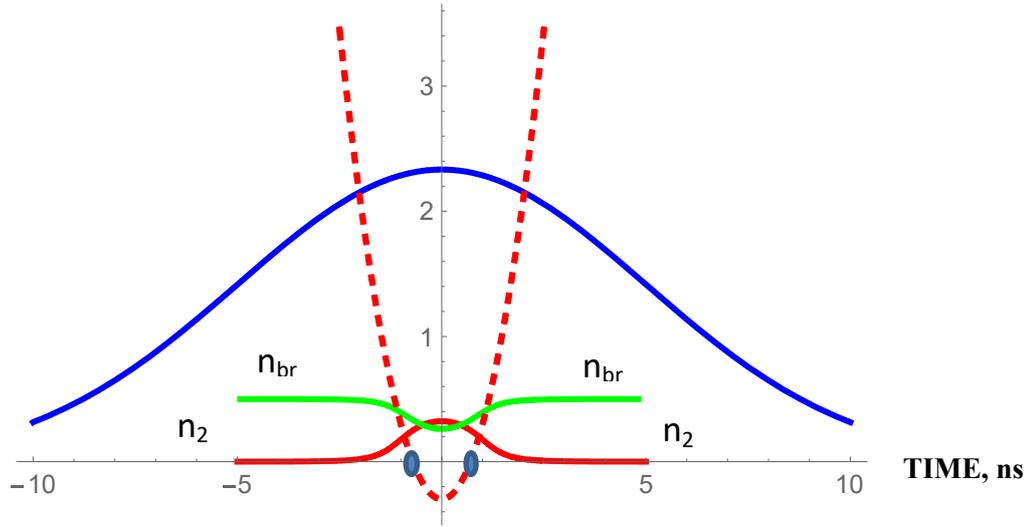

Fig.3. Time dependence of populations of the „bright" superposition (green) and of the excited state (red) in the field of a QFC laser pulse along with the time shape of the laser pulse. (For convinience of presentation the amplitude of the pulse is reduced by 15 times). The frequency detuning $\epsilon(t)$ is shown by dashed line with the two resonance crossing points. Parameters applied are the same as in Fig.2. Note that for equal dipole moments for transitions between the excited state I2> and metastable states I1> and I3>, population of the "bright" state is equal to 0.5.

As can be seen in Fig. 3, resonance with the allowed transitions of the system is reached twice in the case of a quadratic frequency chirp (instead of a single resonance crossing in the case of a linearly chirped pulse), which leads to a return to the ground state of the population of the "bright" component after a temporary population of the excited state.

This effect of a quadratically chirped laser pulse will be used at the next step to create a coherent superposition of metastable states of the considered $\Lambda$- stractured quantum system. The basic physics of the superposition generation is as follows: the phase of the „bright" component after interaction with the laser pulse differs from its initial value (before interaction with the laser pulse) and depends on the parameters of the laser pulse, including its amplitude, duration, detuning of its central frequency and on the speed of the chirp. In the same time, the "dark" superposition component remains untouched when interacting with the laser pulse. As a result, the probability amplitudes of metastable states, and therefore the amplitude and phase of coherence between them, take values that depend on the parameters of the laser pulse.

Using Eq.(5, 9-11) and taking quasinergy $w = w_1$, we obtain for the anplitudes of the „bright" $g_b(t)$ and of the excited state $a_2(t)$ :

$$g_b(t) = \frac{w_1 - \epsilon(t)}{\sqrt{[w_1 - \epsilon(t)]^2 + F(t)^2}} \; g_b(-\infty) \exp[-i\varphi(t)] \;,$$

$$a_2(t) = \frac{F(t)}{\sqrt{[w_1 - \epsilon(t)]^2 + F(t)^2}} \; g_b(-\infty) \exp[-i\varphi(t)]. \qquad (14)$$

where $\varphi(t) = \int_{-\infty}^{t} w_1(t') dt'$ .

For the probability amplitudes $a_{1,3}(t)$ of the ground (metastable) states of our quantum system, we obtain, using equations (3) and (14):

$$a_1(t) = \frac{d_{12}}{d} \frac{w_1 - \epsilon(t)}{\sqrt{[w_1 - \epsilon(t)]^2 + F(t)^2}} g_b(-\infty) \exp[-i\varphi(t)] + C_1,$$

$$a_3(t) = \frac{d_{23}}{d} \frac{w_1 - \epsilon(t)}{\sqrt{[w_1 - \epsilon(t)]^2 + F(t)^2}} g_b(-\infty) \exp[-i\varphi(t)] - C_2,$$

(15)

where $C_1 = \frac{a_1(-\infty)|d_{23}|^2 - a_3(-\infty) d_{12} d_{23}}{d^2}$; $C_2 = \frac{a_1(-\infty) d_{21} d_{23} - a_3(-\infty)|d_{12}|^2}{d^2}$.

At this step, we assume that initially the population of the quantum system under consideration is in one of the metastable states, for example, in the state I1> : $a_1(-\infty) = 1$, $a_3(-\infty) = 0$. Also, for simplicity, we assume that the transition dipole moments are equal: $d_{12} = d_{23}$. These assumptions simplify the expressions (15) significantly: $g_b(-\infty) = g_d(-\infty) = \frac{1}{\sqrt{2}}$ and

$$a_1(t) = \frac{1}{2}[\Gamma(t) \exp[-i\varphi(t)] + 1],$$

$$a_3(t) = \frac{1}{2}[\Gamma(t) \exp[-i\varphi(t)] - 1], \quad (16)$$

where $\Gamma(t) = \frac{w_1 - \epsilon(t)}{\sqrt{[w_1 - \epsilon(t)]^2 + F(t)^2}}$.

Next, we study the dependence of the coherence $\rho_{13}(t) = a_1(t) a_3^*(t)$ between the two metastable states of the quantum system established after action of a QFC laser pulse on the laser pulse parameters, using the results of the dressed states analysis (Eq.16), and direct numerical simulation of the Schrödinger equation (Eq.1).

Assuming a positive speed parameter $\beta > 0$ for a quadratic frequency chirp, (see Fig.3), two different cases are considered below: a) when the central frequency of the laser is less than the resonant frequency of the transition between the metastable and excited states (the detuning is negative at the central frequency of the laser), and b) - when the central frequency of the pulse is greater than the resonant frequency (the detuning is positive at the central frequency of the laser).

These two cases demonstrate two different physical mechanisms of interaction. While in case a) the laser carrier frequency passes through resonance with the transition twice, in case b) the laser frequency has a minimal (positive) detuning at the center of the pulse, without reaching resonance with the transition.

The population value of the "bright" component, temporarily transferred to the excited state, depends differently on the central detuning of the laser pulse in these two cases. In case a), the greater the central (negative) detuning, the greater the part of the "bright" population that temporarily is transferred into the excited state, and the maximum value of the transferred population is equal to the entire initial population of the "bright" component. In case b), the greater the central (positive) detuning, the smaller the part that temporarily goes into the excited state. Note that changing the sign of the chirp speed parameter from $\beta > 0$ to $\beta < 0$ leads to replacement of the situations corresponding to the cases a) and b) considered above.

### 3. Creation of a coherent superposition of the metastable states by a QFC laser pulse

#### 3.1. The case of negative central detuning

We proceed with numerical simulation of the Schrödinger equation (1) for the case of a negative central detuning, as discussed above. The dynamics of the populations of quantum states, as well as the absolute value of the coherence between the metastable states are presented in Fig.4.

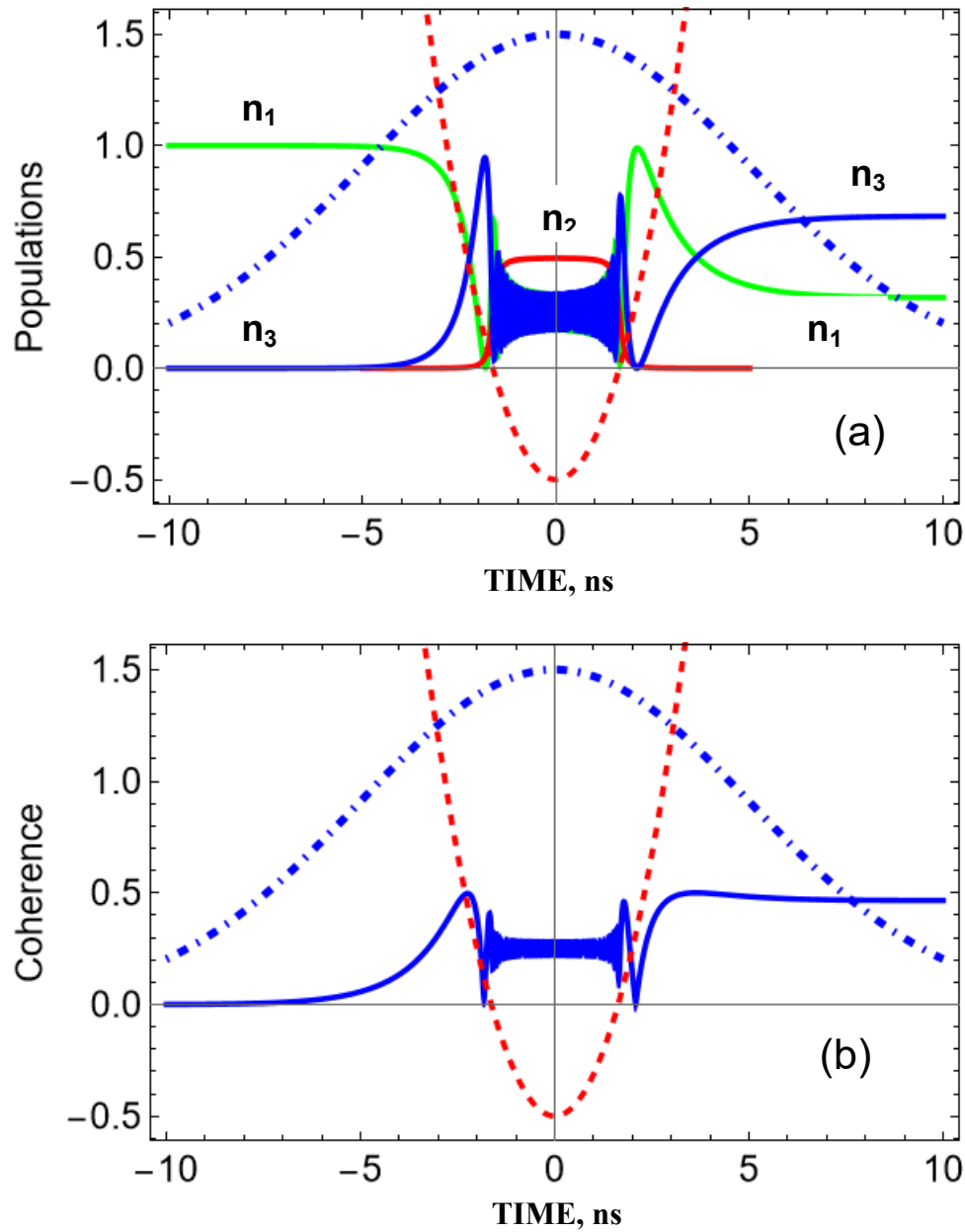

Fig.4. Time dependence (in ns) of the populations of the metastable states |1>, $n_1$ (green), |3>, $n_3$ (blue) and the excited state |2>, $n_2$ (red) – (a); and (b) - the absolute value of the coherence between the metastable states |1> and |3 > in the field of a QFC laser pulse. The system is supposed to be in the state |1> initially. The frequency detuning $\epsilon(t)$ is represented by the dashed line, and the laser pulse shape is represented by the dashed-dotted line. The parameters applied are: the half pulse duration $\tau_p$ = 5 ns, amplitude of the Rabi frequency $w_p$ = 15 GHz, the chirp speed parameter $\beta$ = 25 GHz². For convenience of presentation, the values of the Rabi frequency and laser pulse detuning are reduced by 10 and 400 times, respectively.

The color (density) graph of the absolute value of coherence between two metastable states of the system depending on the detuning of the central frequency and the amplitude of the Rabi frequency is shown in Fig. 5.

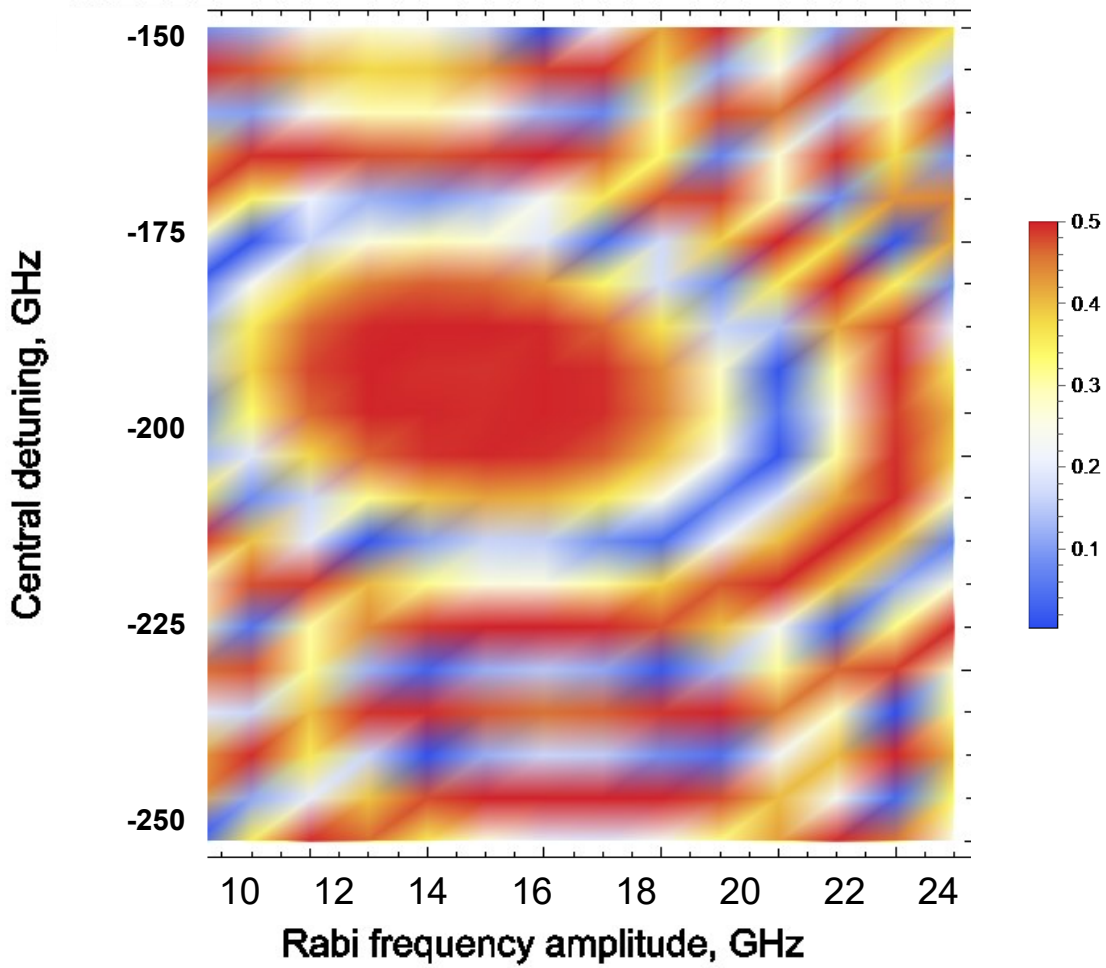

Fig.5. Color (density) plot of the absolute value of coherence established after action of a QFC laser pulse as a function of the negative central detuning and the amplitude of the Rabi frequency. The QFC pulse duration and the chirp speed parameter are the same as in Fig.4.

As follows from Fig. 5, there is a region of QFC pulse parameters where the maximum (near 0.5) coherence can be achieved with extremely high robustness. In addition, from Fig. 5, one can observe a quasi-periodic dependence of the coherence on the central detuning and the Rabi frequency amplitude.

This quasi-periodic dependence can also be seen in Fig. 6 below, where the dependence of the absolute value of the created coherence between metastable states is presented as a function of the (negative) central detuning for a given value of the Rabi frequency amplitude, and in Fig. 7 - depending on the amplitude of the Rabi frequency for a given (negative) value of the central frequency detuning.

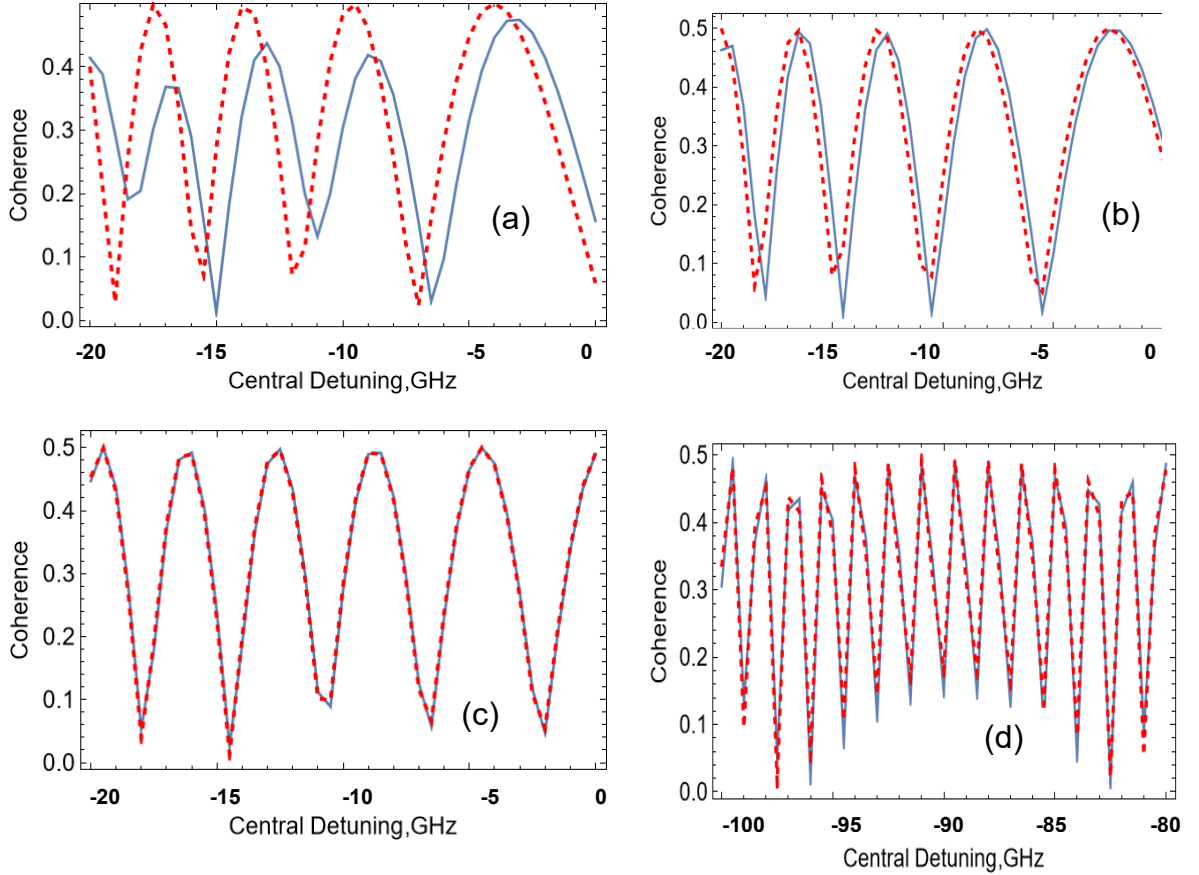

Fig.6. Dependence of the absolute value of coherence between metastable states after action of a QFC laser pulse on the central (negative) detuning $e_{0p}$ for different values of the amplitude $w_p$ of the Rabi frequency of the laser pulse as a result of numerical simulation of the Schrödinger equation (Eq.(1)), (solid lines) and based on the dressed states analysis (Eq.16), dashed lines: (a) - $w_p$ = 1 GHz; (b)- $w_p$ = 5 GHz; (c)- $w_p$ = 15 GHz; (d) - $w_p$ = 15 GHz but for larger negative values of the detuning $e_{0p}$. The QFC pulse duration and the chirp speed parameter are the same as in Fig.5.

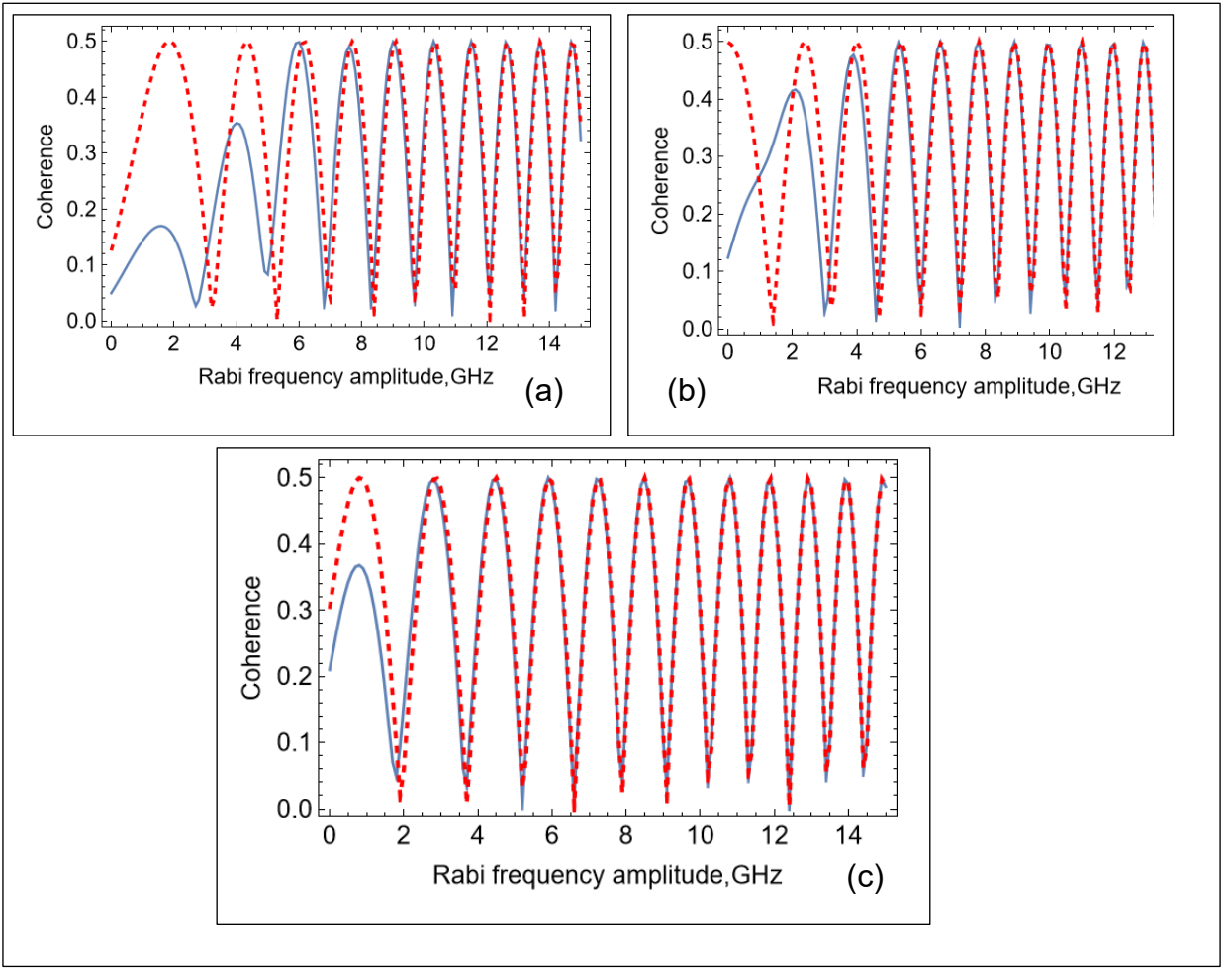

Fig.7. Dependence of the absolute value of coherence between the metastable states after action of a QFC laser pulse as a function of the pulse Rabi frequency amplitude for different values of the central (negative) detuning $e_{0p}$ as a result of numerical simulation of the Schrödinger equation (Eq.(1)), (solid lines), and based on the dressed states analysis (Eq.16), (dashed lines): (a)- $e_{0p}$ = -100 GHz; (b)- $e_{0p}$ = -20 GHz; (c)- $e_{0p}$ = 0. The QFC pulse duration and the chirp speed parameter are the same as in Fig.6.

As can be seen from Figs. 6 and 7, the results based on the numerical solution of the Schrödinger equation are close to the results obtained based on the dressed states approximation, for sufficiently large values of the maximum Rabi amplitudes.

### 3.2. The case of positive central detuning

In this section, we study the effect of a QFC laser pulse on a quantum system with $\Lambda-$structure of operating energy levels under the condition of a positive value of the central frequency detuning (see Fig.8). We perform an analysis similar to the one above but for a positive central frequency detuning. While in the previous case with a negative central frequency detuning the QFC pulse frequency passes through resonance with the system twice, in the case of a positive central detuning, the resonance with the system does not reached at all. This has a number of interesting and useful consequences. One of them, is the negligible excitation of the system, as can be seen in Fig.8(a), (compare with the case of negative central detuning, Fig.4(a)).

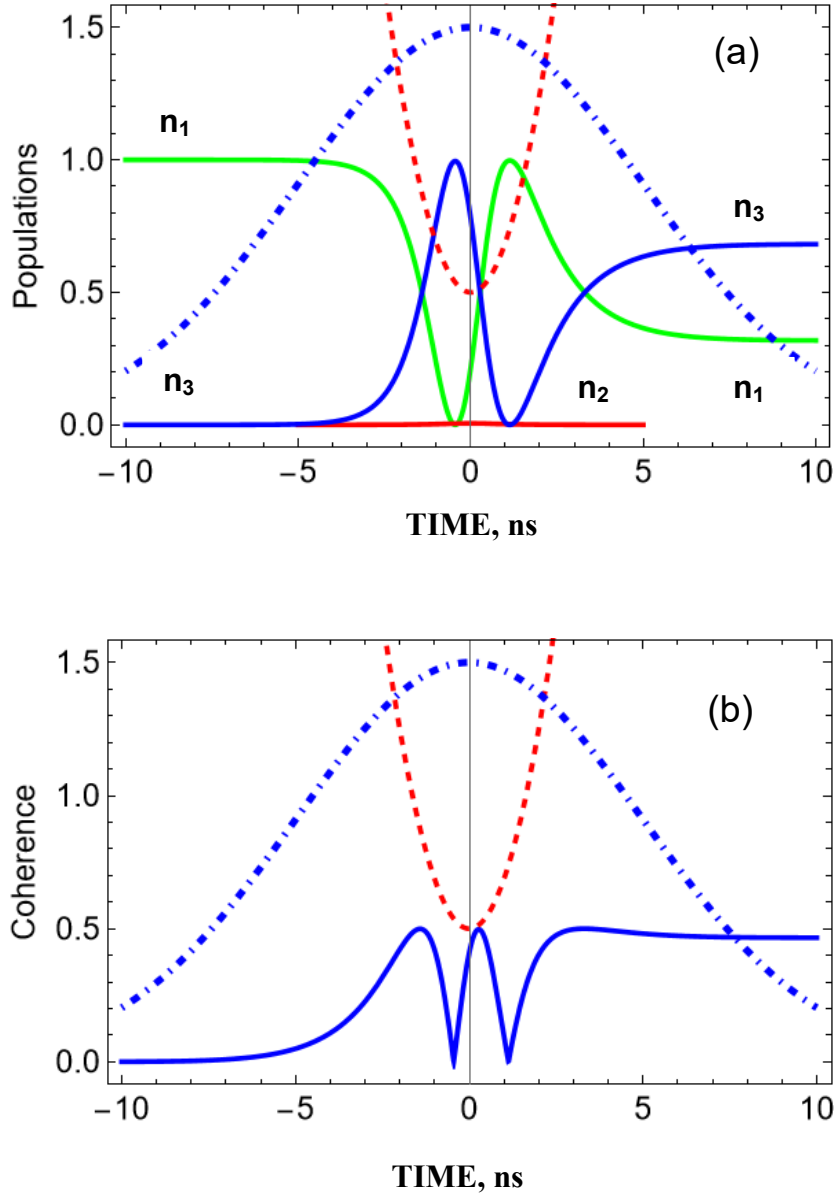

Fig.8. Time dependence of the populations of metastable states |1⟩, $n_1$ (green), and |3⟩, $n_3$ (blue) and the excited state |2⟩, $n_2$ (red) – (a); and (b) - of the absolute value of the coherence between the metastable states |1⟩ and |3⟩ in the field of a QFC laser pulse with positive central frequency detuning. The system is supposed to be in the state |1⟩ initially. The frequency detuning is presented by a dashed line and the laser pulse shape by a dot-dashed line. The parameters applied are: half pulse duration $\tau_p$ = 5 ns, the amplitude of the Rabi frequency $w_p$ = 15 GHz, the chirp speed parameter $\beta$ = 25 GHz$^2$. For convenience of presentation, the values of the Rabi frequency and laser pulse detuning are reduced by 10 and 400 times, respectively.

Below color (density) graphs are presented for dependence of the coherence established after action of a QFC laser pulse on the (positive) central detuning and the amplitude of the Rabi frequency.

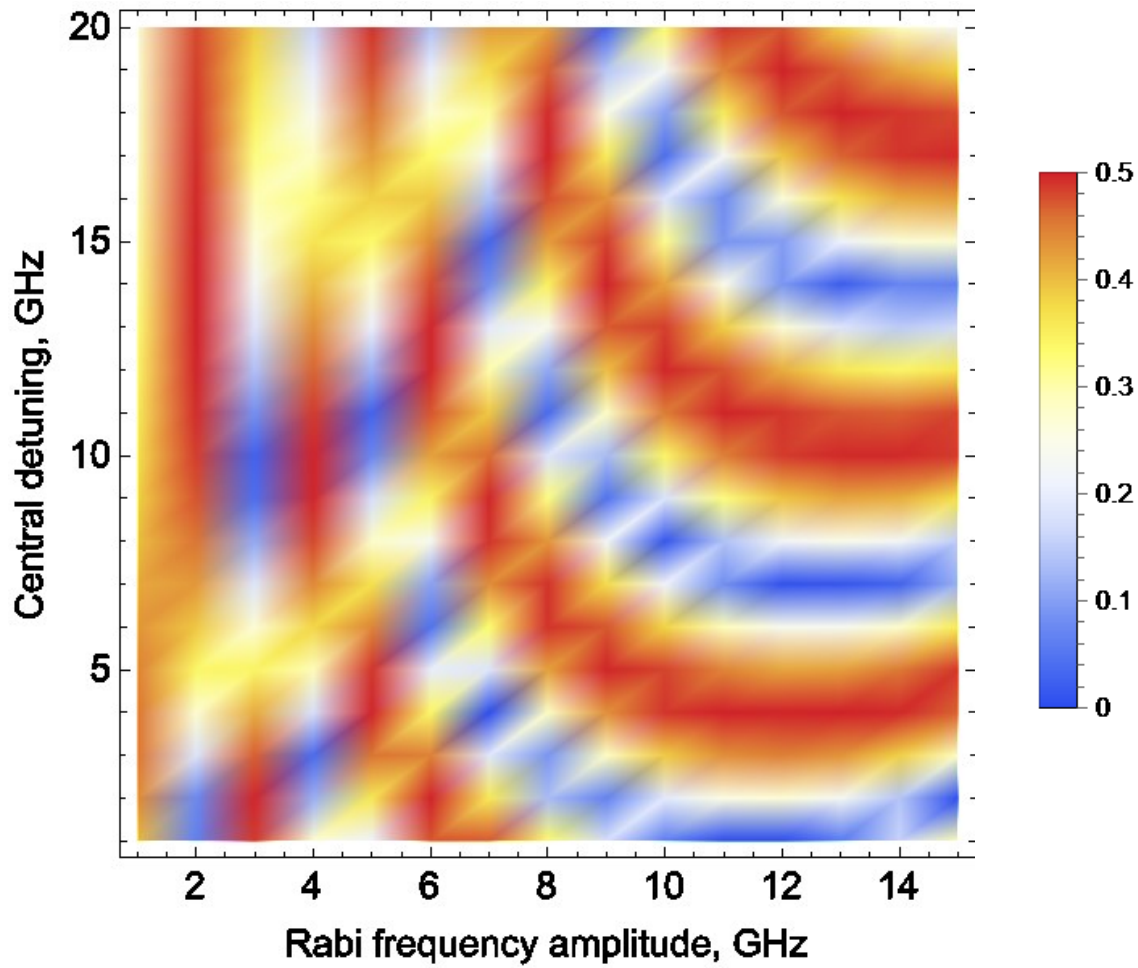

Fig.9. Color (density) plot of the absolute value of the coherence versus *positive* central detuning and Rabi frequency amplitude. The QFC pulse duration and the chirp speed parameter are the same as in Fig.8.

As it follows from Fig.9, there are parameter regions in which the creation of a given coherence may be realized with high robustness. It should be noted that the quasi-periodic dependence of coherence on the central frequency detuning and the Rabi frequency amplitude is similar to the case of negative detuning considered above.
Below in Fig. 10 the dependence of the absolute value of the created coherence between metastable states is presented as a function of the (positive) central detuning for a given value of the Rabi frequency amplitude, and in Fig. 11 as a function of the Rabi frequency amplitude for a given (positive) value of the central frequency detuning.

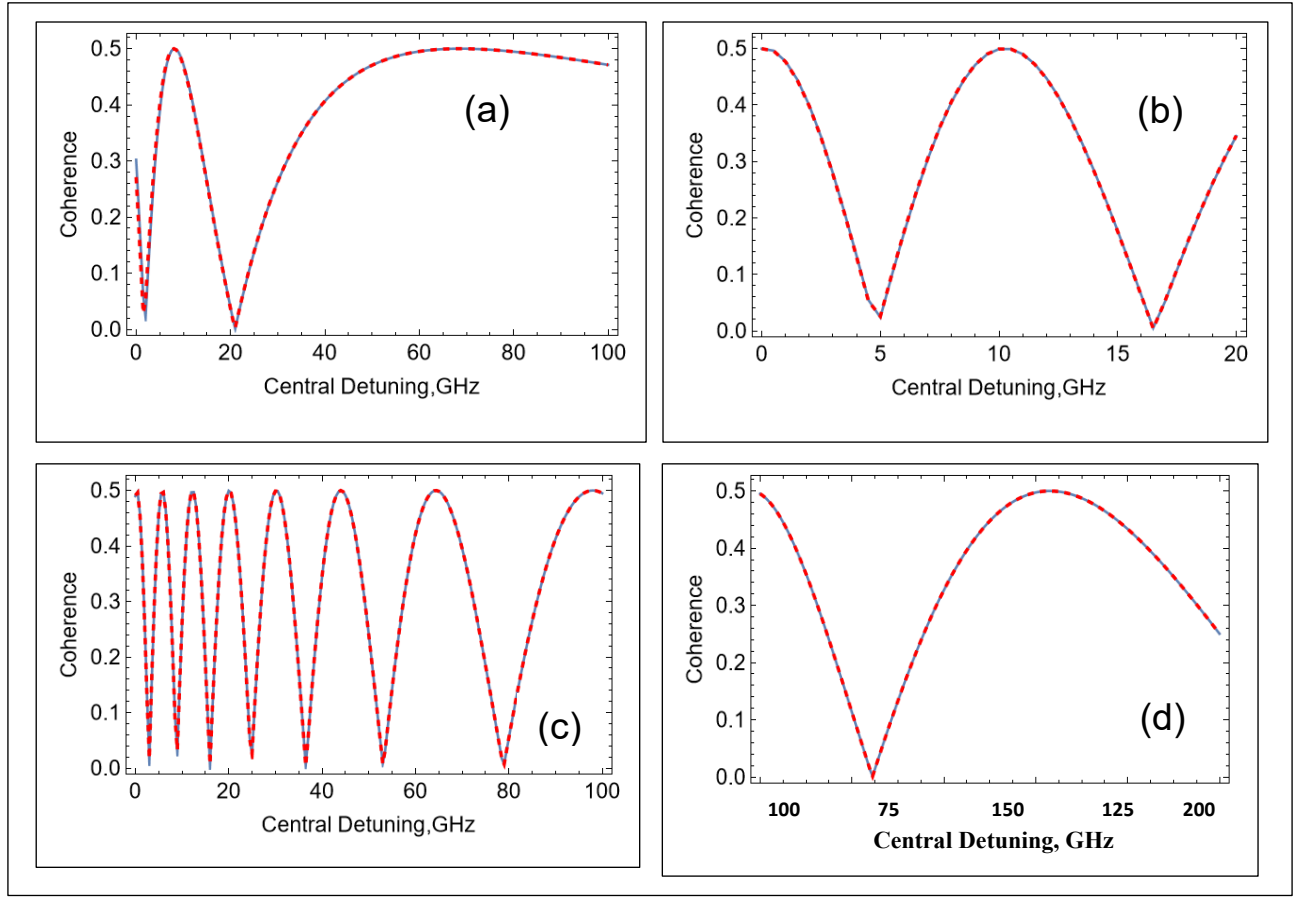

Fig.10. Dependence of the absolute value of the coherence between the metastable states after action of a QFC laser pulse on the central (positive) detuning $e_{p0}$ for different values of the Rabi frequency amplitude $w_{p0}$ as a result of numerical simulation of the Schrödinger equation (Eq.(1)), (solid lines) and based on the dressed states analysis (Eq.16), dashed lines: (a) - $w_p$ = 5 GHz; (b)- $w_p$ =15 GHz; (c)- $w_p$ = 15 GHz with a larger range of variation of the detuning $e_{p0}$; (d) - $w_p$ = 15 GHz but for even greater positive values of the detuning $e_{0p}$. The QFC pulse duration and the chirp speed parameter are the same as in Fig.9.

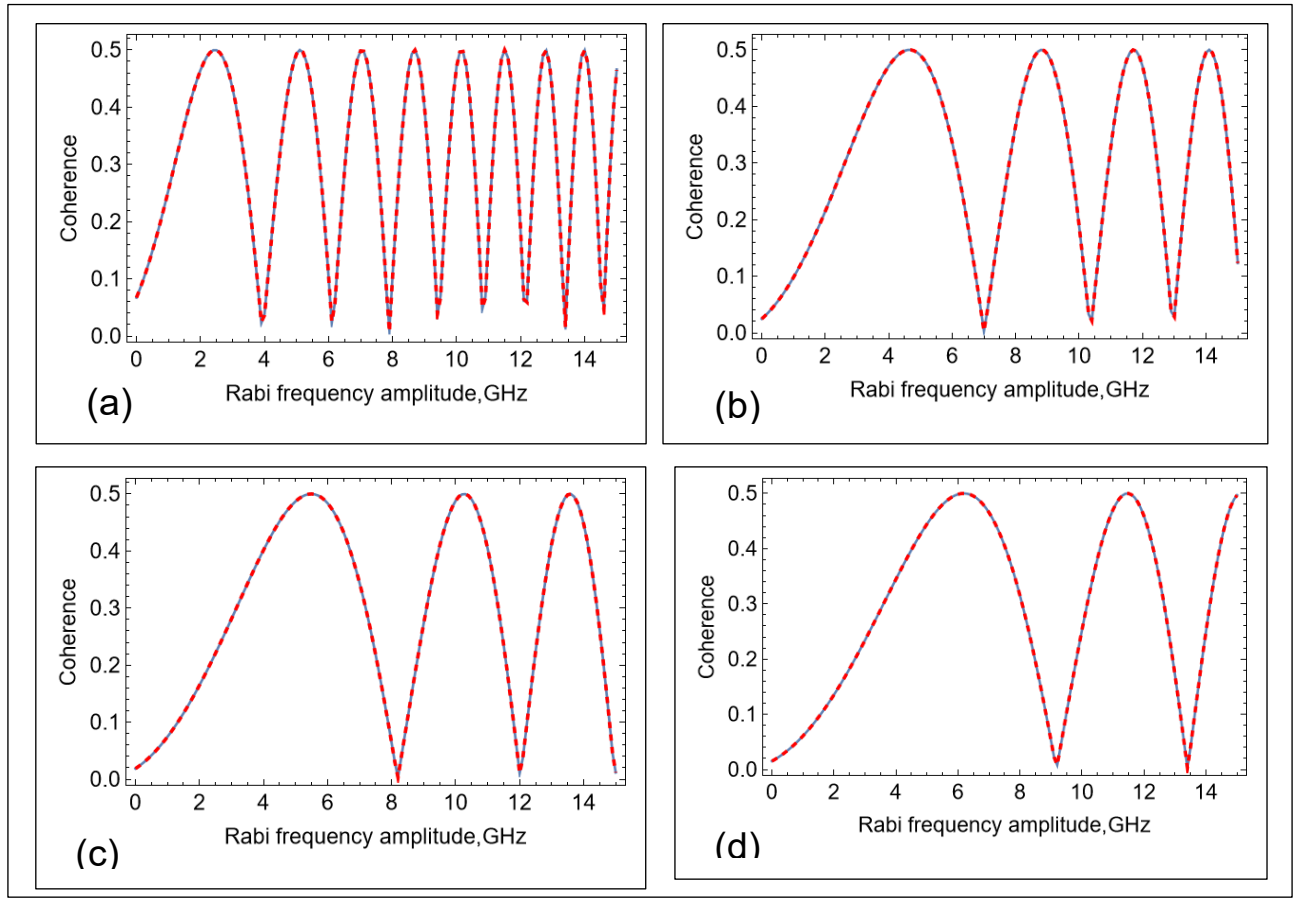

Fig.11. Dependence of the absolute value of the coherence between the metastable states after action of a QFC laser pulse on the Rabi frequency amplitude for different values of the central (positive) detuning $e_{p0}$ as a result of numerical simulation of the Schrödinger equation (Eq.(1)), (solid lines), and based on the dressed states analysis (Eq.16), (dashed lines): (a) – $e_{0p}$ = 20 GHz; (b)- $e_{0p}$ = 100 GHz; (c)- $e_{0p}$ = 150 GHz, d)- $e_{0p}$ = 200 GHz. The QFC pulse duration and the chirp speed parameter are the same as in Fig.10.

As it can be seen from Figs.10 and 11, the solutions based on the solution of the Schrödinger equation (Eq.1) and those obtained on the basis of the dressed states approximation (Eq.(16)) coincide with each other in the considered parameters range.

### 4. Conclusions

In conclusion, interaction of a QFC laser pulse with a quantum system having $\Lambda$-configuration of working energy levels has been studied with the aim of creating arbitrary values of coherence between metastable states of the system with negligibly small and temporary excitation of the system. Two cases of central frequency detuning have been considered: negative, when the laser frequency passes through resonance with the system twice, and positive, when the laser pulse frequency approaches resonance without crossing it. It is shown that in both cases it is possible to create an arbitrary coherence value by varying the parameters of the laser pulse, including its central frequency and amplitude.
In both cases considered, there are ranges of variation of these parameters where creation of the coherence is robust against small to medium variation of the laser pulse parameters. In both cases, the absolute value of the created coherence is a quasi-periodic function of the central frequency and the (Rabi frequency) amplitude of the laser pulse.

However, the case of positive central detuning seems more preferable due to the insignificant temporary excitation of the system, as well as due to the larger values of the (quasi-) periods of coherence variation when changing the parameters of the laser pulse, making its creation more reliable and predictable.

It is worth noting that the quadratic frequency chirp of the laser pulse considered in this paper is just a simplification of the non-monotonic temporal behavior of the bell-shaped (or inverse bell-shaped) time-dependent frequency chirp. Thus, all the effects of creating coherence between metastable states are also valid for more general, bell-shaped frequency chirped laser pulses.

It should be noted that the coherence creation scheme with QCF laser pulses discussed in this manuscript may be generalized to spatially extend optically thick media taking into account propagation of laser fields. This will be the topic of our next article.

**References**

1. M. Kasevich and S. Chu, Atomic interferometry using stimulated Raman transitions. Phys. Rev. Lett. **67**, 181-184 (1991);
2. M. Weitz, B. C. Young, and S. Chu, Atomic interferometer based on adiabatic population transfer. Phys. Rev. Lett.. **73**, 2563-2566 (1994).
3. J. Lawall and M. Prentiss, Demonstration of a novel atomic beam splitter
Phys. Rev. Lett. **72**, 993-997 (1994).
4. D. L. Butts, J. M. Kinast, K. Kotru, A. M. Radojevic, B. P. Timmons and R. E. Stoner, Coherent population trapping in Raman-pulse atom interferometry Phys. Rev. A
84 043613 (2011).
5. H. Theuer, R. Unanyan, C. Habscheid, K. Klein, and K. Bergmann. Novel laser controlled variable matter wave beamsplitter, Opt. Express **4**, 77–83 (1999).
6. P. Brumer and M. Shapiro, "Laser control of molecular processes," Annu. Rev. Phys. Chem. **43**, 257–282 (1992).
7. X. Li and G. A. Parker. Theory of laser enhancement of ultracold reactions, J. Chem. Phys. 128 184113 (2008).
8. S. Chatterjee and S. S. Bhattacharyya. Selective excitation of LI2 by chirped laser pulses with all possible interstate radiative couplings J. Chem. Phys. 133 164313 (2010).
9. C. P. Williams and S. H. Clearwater, *Explorations in Quantum Computing* (Springer-Verlag,1997).
10. D. Bouwmeester, A. Ekert, and A. Zeilinger, *The Physics of Quantum Information: Quantum Cryptography, Quantum Teleportation, Quantum Computation* (Springer-Verlag, 2000).
11. P. Kaufmann, T. F. Gloger, D. Kaufmann, M. Johanning and C.Wunderlich. High-fidelity preservation of quantum information during trapped-ion transport Phys. Rev. Lett. 120 010501 (2018).
12. M. G. Bason, M. Viteau, N. Maloss, P. Huillery, E. Arimondo, D. Ciampini,
R. Fazio, V. Giovannetti, R. Mannella and O. Morsch. High-fidelity quantum driving Nat. Phys. 8:147 (2012).
13. M. Saffman, T.G. Walker and K. Mølmer. Quantum information with Rydberg atoms Rev. Mod. Phys. 82 2313–63 (2010).
14. C. H. Bennett and D. P. DiVincenzo. Quantum information and computation Nature 404, 247 (2000).
15. A. S. Parkins, P. Marte, P. Zoller, and H. J. Kimble, Synthesis of arbitrary quantum states via adiabatic transfer of Zeeman coherence, Phys. Rev. Lett. **71**, 3095–3098 (1993).

**16.** G. P. Djotyan, J. S. Bakos, G. Demeter, P. N. Ignácz, M. A. Kedves, Zs. Sörlei, J. Szigeti, and Z. L. Tóth, Coherent population transfer in Rb atoms by frequency-chirped laser pulses. Phys. Rev. A 68, 053409 (2003).
17. G. Liu, V. Zakharov, T. Collins, P. Gould, and S. A. Malinovskaya, Population inversion in hyperfine states of Rb, with a single nanosecond chirped pulse in the framework of a four-level system, Phys. Rev. A 89, 041803(R) (2014).
18. K. Varga-Umbrich, J.S. Bakos, G. P. Djotyan, Zs. Sörlei, G. Demeter, P. N. Ignacz, B. Ráczkevi, J. Szigeti1, and M.A. Kedves, Coherent manipulation of trapped Rb atoms by overlapping frequency-chirped laser pulses: theory and experiment. Eur. Phys. J. D (2022) 76 :70.
19. F. Vewinger, M. Heinz, R. G. Fernandez, N. V. Vitanov, and K. Bergmann. Creation and measurement of a coherent superposition of quantum states, Phys. Rev. Lett. **91**, 213001 (2003).
20. N. Sangouard, S. Guérin, L. P. Yatsenko, and T. Halfmann, Preparation of coherent superposition in a three-state system by adiabatic passage, Phys. Rev. A **70**, 013415 (2004).
21. L. Yatsenko, N. V. Vitanov, B. W. Shore, T. Rickes, and K. Bergmann. Creation of coherent superpositions using Stark-chirped rapid adiabatic passage, Opt. Commun. **204**, 413–423 (2002).
22. G. Djotyan, J. Bakos, Zs. Sörlei, J. Szigeti, P. Ignacz, and Z. Toth, Interaction of a sequence of frequency-chirped bichromatic laser pulses with an ensemble of lambda- atoms: population trapping and coherent optical pumping, Laser Phys. **10**, 355–359 (2000).
23. G. P. Djotyan, J. S. Bakos, Zs. Sörlei, and J. Szigeti. Coherent control of atomic quantum states by single frequency-chirped laser pulses. Phys. Rev. A 70, 063406 (2004).
24. Z Zhang, X Yang, and X Yan, Population transfer and generation of arbitrary superpositions of quantum states in a four-level system using a single-chirped laser pulse, J. Opt. Soc. Am. B, Vol. 30, pp. 1017 – 1021 ( 2013 ).
25**.** G. P. Djotyan, J. S. Bakos, G. Demeter, Zs. Sörlei, J. Szigeti, and D. Dzsotjan. Creation of a coherent superposition of quantum states by a single frequency-chirped short laser pulse. J. Opt. Soc. Am. B., **25**, 166-174 (2008).
26. G. P. Djotyan, N. Sandor, J. S. Bakos, Zs. Sörlei, An extremely robust strong-field control of atomic coherence, Opt. Express, **19**, 17493-17499 (2011).
27. M. Ndong, G.P. Djotyan, A Ruschhaupt and S Guérin, Robust coherent superposition of states by single-shot shaped pulse, J. Phys. B: At. Mol. Opt. Phys. **48**. 174007, (2015).
28. J. B. Watson, A. Saprera, X. Chen, and K. Burnett, Harmonic generation from a coherent superposition of states, Phys. Rev. A **53**, R1962–R1965 (1996).
29. M. Jain, Hui Xia, G. Y. Yin, A. J. Merriam, and S. E. Harris, Efficient nonlinear frequency conversion with maximal atomic coherence, Phys. Rev. Lett. **77**, 4326–4329 (1996).
30. A. V. Sokolov, D. R. Walker, D. D. Yavuz, G. Y. Yin, and S. F. Harris, Raman generation by phased and antiphased molecular states, Phys. Rev. Lett. **85**, 562–565 (2000).
31. M. D. Lukin, P. R. Hemmer, M. Loeffler, and M. Scully, Resonant enhancement of parametric processes via radiative interference and induced coherence, Phys. Rev. Lett. **81**, 2675–2678 (1998).
32. E. Korsunsky, T. Halfmann, J. P. Marangos, M. Fleischhauer, and K. Bergmann, Analytical study of fourwave mixing with large atomic coherence, Eur. Phys. J. D **23**, 167–180 (2003).
33. T. Rickes, J. P. Marangos, and T. Halfmann, Enhancement of third-harmonic generation by Stark-chirped rapid adiabatic passage, Opt. Commun. **227**, 133–142 (2003).
34. J T Bahns, W C Stwalley and P L Gould Laser cooling of molecules: a sequential scheme for rotation, translation, and vibration J. Chem. Phys. **104**, 9689–97 (1996).
35. M Dupont-Nivet, M Casiulis, T Laudat, C I Westbrook and S Schwartz Microwave-stimulated Raman adiabatic passage in a Bose–Einstein condensate on an atom chip Phys. Rev. A 91 053420 (2015).

36. G. P. Djotyan, J. S. Bakos, and Zs. Sörlei. Coherent writing and reading of information using frequency-chirped short bichromatic laser pulses, Opt. Express **4**, 113–120 (1999).
37. G. P. Djotyan, J. S. Bakos, and Zs. Sörlei, Three-level Λ atom in the field of frequency-chirped bichromatic laser pulses: Writing and storage of optical phase information. Phys. Rev. A, **64**, 013408 (2001).
38. O.V. Ivakhnenko, S.N. Shevchenko, F. Nori, Nonadiabatic Landau–Zener Stückelberg–Majorana transitions, dynamics, and interference, Physics Reports **995,** 1–89 (2023)
39. K. Bergmann, H. Theuer, and B. W. Shore. Coherent population transfer among quantum states of atoms and molecules, Rev. Mod. Phys. **70**, 1003–1025 (1998).
40. Vitanov N V, Rangelov A A, Shore B W and Bergmann K 2017 Stimulated Raman adiabatic passage in physics, chemistry, and beyond Rev. Mod. Phys. 89 015006,
41. N. V. Vitanov, T. Halfmann, B. W. Shore, and K. Bergmann, Annu. Rev. Phys. Chem. **52**, 763 (2001).
42. V S Malinovsky and J L Krause General theory of population transfer by adiabatic rapid passage with intense, chirped laser pulses Eur. Phys. J. D **14**, 147–55 (2001).
43. P. Král, I. Thanopulos and M. Shapiro. Colloquium: coherently controlled adiabatic passage Rev. Mod. Phys. 79 53–77 (2007).
44. C. Ding, R. Yu, X. Hao, D. Zhang. Controllable population dynamics in Landau-quantized graphene. Scientific Reports, **8,** 1530 (2018).
45. F. Kappe, Y. Karli, G. Wilbur, R. G. Krämer, S. Ghosh, R. Schwarz, M. Kaiser, T. K. Bracht, D. E. Reiter, S. Nolte, K. C. Hall, G. Weihs, V. Remesh. Chirped Pulses Meet Quantum Dots: Innovations, Challenges, and Future Perspectives. Advanced Quantum Technologies, 8, 2300352 (2025).
46. A. P. Djotyan, G. P. Djotyan, and A. A. Avetisyan. Coherent Control of Lowest States of a Shallow Impurity in Graphene Monolayer by Phase Modulated Laser Pulses. Journal of Contemporary Physics, 59, 272–278 (2024).
47. N.Sandor, J. S. Bakos, Zs. Sörlei, and G. P. Djotyan. Creation of coherent superposition states in inhomogeneously broadened media with relaxation JOSA B, **28**, 2785-2796 (2011).
48. C. Xu, A. Poudel and M. G. Vavilov, Nonadiabatic dynamics of a slowly driven dissipative two-level system. Phys. Rev. A **89**, 052102 (2014).
49. Levy A, Torrontegui E and Kosloff R, Action-noise-assisted quantum control. Phys. Rev. A **96**, 033417 (2017).
50. A. A. Avetisyan, A. P. Djotyan, G. P. Djotyan, A. L. Vartanian, and A. L. Asatryan. Coherent control of a shallow impurity quantum states in graphene monolayer by bichromatic phase modulated laser pulses: Influence of relaxation processes. Journal of Contemporary Physics, **60**, 51–61 (2025).
51. F. Ahmadinouri, M Hosseini and F Sarreshtedari, Investigation of robust population transfer using quadratically chirped laser interacting with a two-level system, Phys. Scr. **94,** 105404 (2019).
52. P. Meystre and M. Sargent III, *Elements of Quantum Optics* (Springer-Verlag, New York, 1991).
53. L. Allen and J. H. Eberly, *Optical Resonance and Two-Level Atoms* (Dover, New York, 1987).